# Lithium and sodium storage on tetracyanoethylene (TCNE) and TCNE-(doped)-graphene complexes: a computational study

Yingqian Chen and Sergei Manzhos*

Department of Mechanical Engineering, National University of Singapore, Block EA #07-08, 9 Engineering Drive 1, Singapore 117575, Singapore

*E-mail: mpemanzh@nus.edu.sg; Tel: +65 6516 4605

## Abstract

Li and Na attachment to free tetracyanoethylene (TCNE) molecules and TCNE adsorbed on doped graphene is studied using density functional theory. While TCNE is adsorbed only weakly on ideal graphene, we identified a configuration in which TCNE is chemisorbed on Al-doped graphene via molecule's C atom and a surface oxygen atom. Up to four (five) Li and Na atoms can be stored on both free (adsorbed) TCNE with binding energies stronger than cohesive energies of the Li and Na metals. When storing up to three atoms per molecule, it should be possible to avoid reduction of common battery electrolytes. TCNE immobilized on a conducting graphene-based substrate could therefore become an efficient anode material for organic Li and Na ion batteries. Importantly, there is no significant difference either in specific capacity (per unit mass of material excluding Li/Na) nor in voltage between Li and Na storage, which makes this kind of approach very promising for post-Li storage in general.

## 1. Introduction

With the growing use of renewable but intermittent sources of electricity (such as wind and solar) and with the unrelenting advance of all-electric vehicles, demand for energy storage and, specifically, electrochemical storage is growing [1]. Lithium ion batteries provide today the highest energy density and cycle rate and life among commercial battery technologies [2]. Those, however, still require improvements [3]. Especially for efficient grid storage, to achieve load balancing and possibly price arbitrage, storage on massive scale, allowing for very rapid (within minutes) charge and discharge, is needed [4, 5]. Such rates of charge-discharge are beyond those of present commercial Li ion batteries [3]. Electrode materials of available Li ion batteries often include expensive or poisonous components like Co (for example, 30% of global Co production is consumed by Li ion batteries [6]). Sodium ion batteries, on the other hand, are the most promising candidate technology for bulk electrochemical storage. Lithium deposits are geographically concentrated, and Li might become too expensive to use Li ion batteries on a large scale [7, 8], while sodium is abundant and cheap. However, it appears that it is more difficult to make (inorganic) electrode materials for Na storage with suitable thermodynamics and kinetics compared to materials for Li storage [9-12].

Organic electrodes are a way to achieve high rate (high power) and environment-friendly batteries. In recent years, exciting progress in the development of organic electrodes has been made [13]. Capacities of up to 900 mAh/g (i.e. competitive with inorganic electrodes) and rates of up to 1000C (unprecedented for inorganic electrodes) have been reported [13]. Moreover, organic electrodes are also promising for post-Li storage [14, 15], which will have to be developed for making massive deployment of electrochemical batteries feasible [16]. Organic electrodes can be made from common feedstock (e.g. biomass) and are therefore sustainable. To develop practical (commercial) organic batteries, many issues still need to be resolved. Specifically, stability of organic materials (including dissolution into the electrolytes) and voltage control are needed to realize organic cathodes and anodes and to maximize power and energy density. Also, electrical conductivity needs to be improved [13]. We note that the mechanism of Li storage in reported crystalline organic materials is still not fully understood [17]. Those can suffer from slow kinetics due to phase transitions induced by cycling and from lower capacity due to the impossibility to use all Li storage sites of the constituent molecule in an organic crystal [18]. To maximize specific capacity, multiple Li or Na atoms need to be stored per formula unit. Li storage on molecular (as opposed to organic crystals) materials has the potential to achieve high rate and high capacity. The problem is that molecules need to be immobilized and to remain connected to the electrode during cycling.

Here, we present a density functional theory (DFT) computational study of the possibilities of Li and Na storage on tetracyanoethylene (TCNE). This is a light molecule which is expected to attach up to 4 Li atoms [19], which would result in a theoretical specific capacity of about 840 $mAh/g_{(TCNE)}$. We confirm that up to 4 Li or Na atoms can attach to a TCNE molecule with a binding energy exceeding the cohesive energy of Li or Na metal. As a way to provide electrode stability and electrical connectivity, we study anchoring TCNE on graphene. Graphene is a substrate of choice due to its high surface area and electronic as well as ionic ($Li^+$ diffusion) conductance [20-22]. We find that TCNE interacts only weakly with ideal graphene, consistent with the previously reported weak interaction between tetracyanides and graphene [23]. We therefore identify an adsorption configuration on Al-doped graphene with strong binding [24]. We then confirm that not only all Li and Na adsorption sites of the TCNE molecule are preserved on the adsorbed TCNE, up to five Li or Na atoms can be stored on it. Also, Li and Na adsorption energies are somewhat stronger than that on free TCNE. We conclude that TCNE immobilized on a conducting graphene-based substrate could become an efficient anode material for organic Li and Na ion batteries.

## 2. Methods

Molecular and electronic structures were optimized with DFT [25] using the SIESTA code [26]. The PBE exchange-correlation functional[27] and a DZP (double-ζ polarized) basis set were used. The basis set was optimized to reproduce cohesive energies of C, Al, Li and Na [28, 29]. For other elements, the basis was generated with the setting “PAO.EnergyShift = 0.01 Ry”. Geometries were optimized until forces on all atoms were below 0.03 eV/Å. A cutoff of 100 Ry was used for the Fourier expansion of the density, and oversampling of the Fourier grid was used to minimize the eggbox effect. An electronic temperature of 500 K was used to speed up the convergence. A ~15x17 Å graphene sheet was used with the vacuum layer of about 20 Å. Due to the large size of the supercell, the Brillouin zone was sampled at the 2×2×1 Monkhorst-Pack points [30]. The calculations were done for TCNE-(doped)-graphene as well as Li/Na-TCNE complexes and Li/Na-TCNE-O-Al-graphene complexes. Calculations with a free TCNE molecule were computed in the same supercell as the graphene-containing systems (at the Γ point). Spin polarization was used in all calculations. Charge transfer between Li/Na atoms and free TCNE or absorbed TCNE was analyzed using Mulliken charges as well as charge density difference maps.

The binding energy ($E_b$) per Li/Na atom was computed as

$$E_b=(E_{n\mathrm{X}/sys}-E_{sys}-nE_{\mathrm{X}})/n, \quad (1)$$

where $E_{n\mathrm{X}/sys}$ is the total energy of $n$ X atoms attached to *sys*, where *sys* is TCNE or TCNE-O-Al-graphene and X = Li or Na; $E_{sys}$ is the total energy of *sys*, and $E_{\mathrm{X}}$ is the total energy of an X atom in a vacuum box (of the same size as the supercell). The binding energies were corrected for the BSSE (basis set superposition error) by using a counterpoise correction [31].

We also computed binding energies including the vdW correction by Grimme [32] (DFT-D) with parameters taken from Ref. [32]. With DFT-D, force tolerance was 0.04 eV/ Å. While GGA DFT is known to underestimate vdW bonding, the DFT-D, as a semi-empirical approach, makes the calculation non- ab initio and may overestimate bonding. Proper description of TCNE-graphene interaction should require DFT-D; however, even though DFT-D with Grimme's correction has been used recently for studies of Li interaction with graphene-like materials [33-35], it is not expected to be reliable for systems with charge donation, which the present systems are (as is also the Li/Na-graphene system) due a significant iconicity of the Li/Na-TCNE bond (as shown below). We therefore provide numbers computed by both DFT and DFT-D to show that both approaches lead to similar conclusions. The same amount of BSSE correction as in DFT was used for DFT-D calculations (due to similar geometries resulting from DFT and DFT-D; we verified that the amount of BSSE correction changes by about 0.01 eV between DFT and DFT-D calculations while the correction itself is about 0.2 eV).

Bond formation between Li/Na atoms and *sys* was analyzed using charge density difference which was computed as

$$\Delta\rho = \rho_{n\mathrm{X}/sys} - \rho_{sys} - \rho_{n\mathrm{X}}, \qquad (2)$$

where all charge densities ρ are computed at the atomic positions of *n*X/*sys* complexes.

## 3. Results

For all systems, we studied all combinations of adsorption sites of one to five Li/Na atoms. Only stable non-equivalent configurations identified with both DFT and DFT-D are shown below. Figure 1 shows Li/Na complexes with a TCNE molecule and lists corresponding binding energies. A free TCNE molecule can store up to four Li or Na atoms. The computed cohesive energy of Li metal is $E_{coh}^{Li}$ = -1.67 (-1.78) eV (here and below, the numbers in parentheses are for DFT-D) and $E_{coh}^{Na}$ = -1.14 (-1.31) eV [28, 29] (Non-BSSE corrected values are used for $E_{coh}$, as the basis sets of Na and Li were tuned to reproduce their respective $E_{coh}$). Therefore, an isolated TCNE molecule can operate as an anode with "voltages" ($V = E_{coh} - E_b$) in the range of 0.29-1.58 (0.23-1.67) V for Li storage and 0.38-1.60 (0.36-1.61) V for Na storage with the final state of charge 4Li/Na-TCNE, corresponding to a specific capacity of about 840 mAh/$g_{(TCNE)}$.

Next, we studied possible immobilization of TCNE on graphene (Figure 2). The binding energy of TCNE on pristine graphene is -0.24 (-1.16) eV and corresponds to physisorption, which is not expected to provide either stability or electric connectivity of the electrode. On Al-doped graphene (Figure 2, middle), $E_b$ = -0.92 (-1.76) eV – slightly stronger. The strongest bonding is achieved by adsorbing TCNE via its C atom and a surface oxygen atom [36] (Figure 2, right), $E_b$ = -2.08 (-2.80) eV. This configuration provides a strong chemical bonding and the accessibility of all Li adsorption sites of a free TCNE molecule.

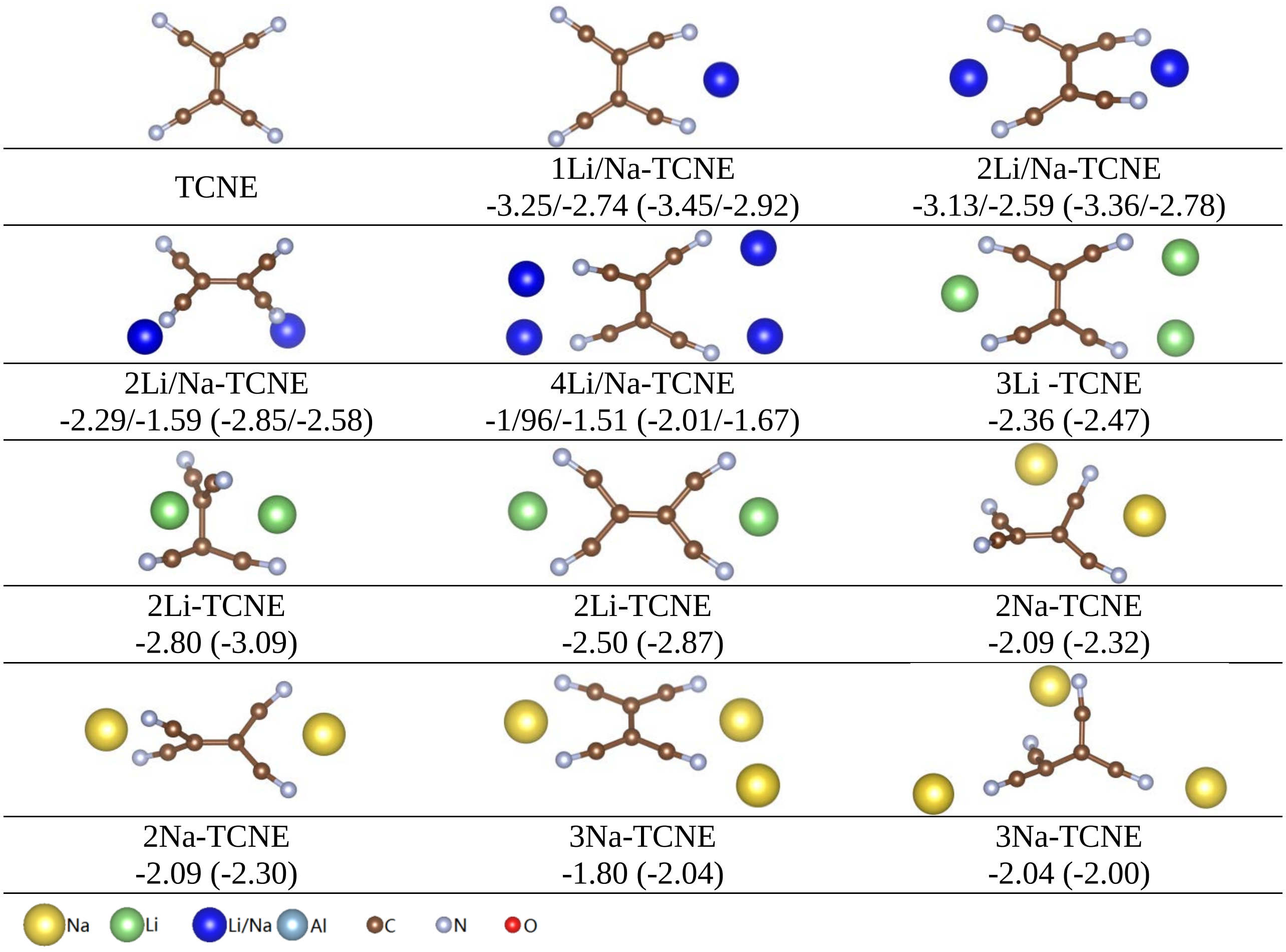

**Figure 1.** Complexes of 1-4 Li/Na atoms with a free TCNE molecule and corresponding binding energies (per Li/Na atom) $E_b$, in eV. Here and elsewhere, DFT-D values are given in parentheses. Configurations identical for Li and Na are shown with blue atoms labelled “Li/Na”.

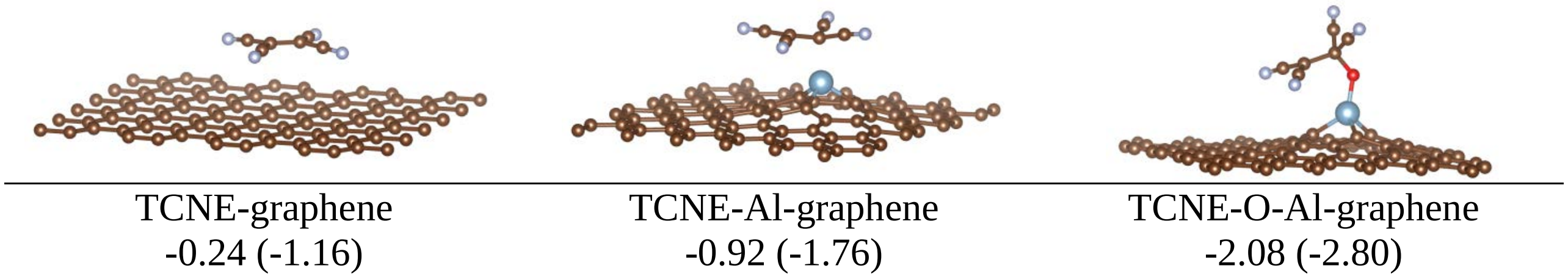

**Figure 2.** Configurations of TCNE adsorbed on pristine and Al-doped graphene and corresponding binding energies $E_b$, in eV, obtained with DFT (DFT-D). Atom color scheme is the same as in Figure 1.

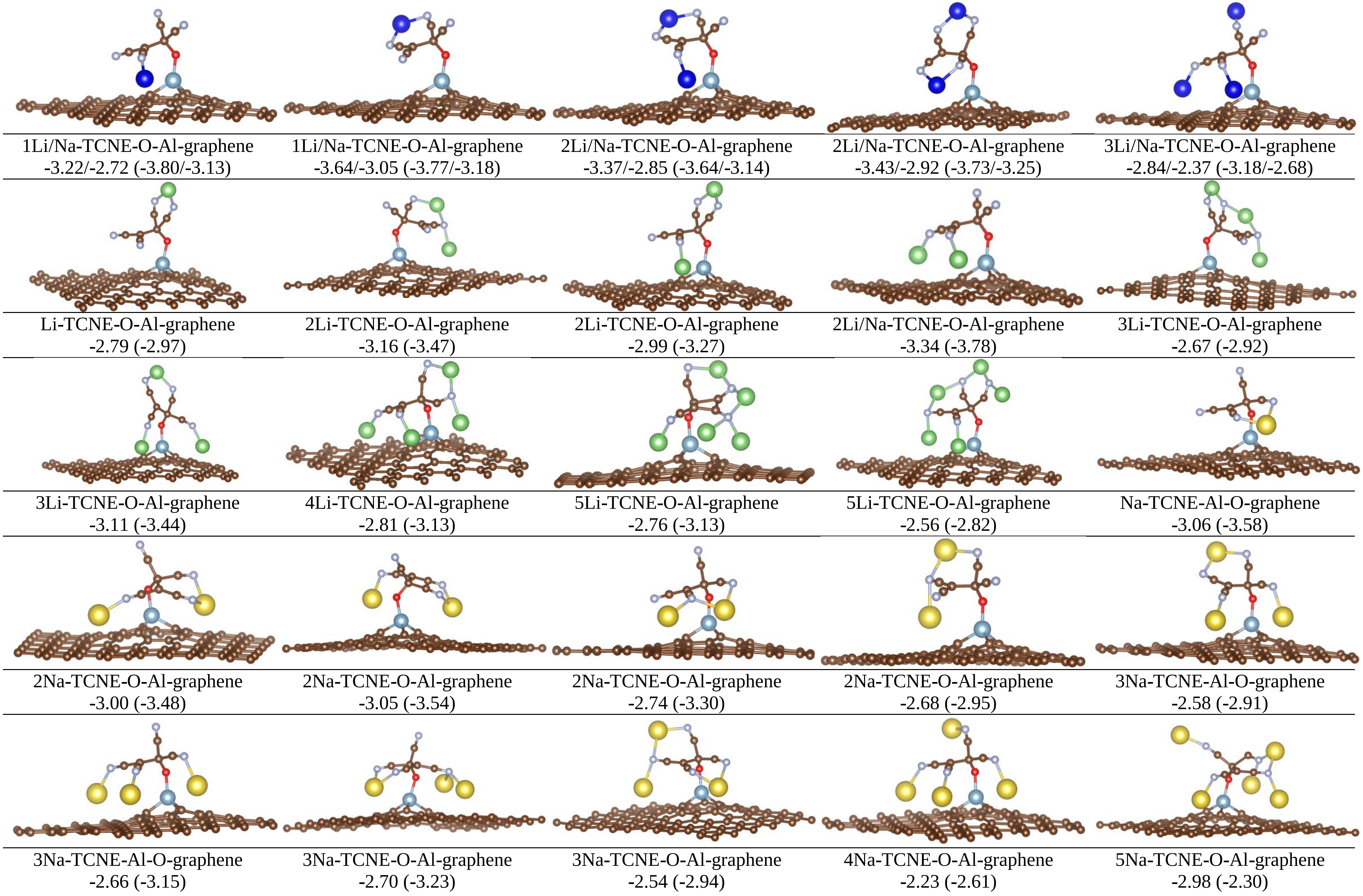

**Figure 3.** Complexes of 1-5 Li/Na atoms with TCNE-O-Al-graphene and corresponding binding energies $E_b$, in eV (Li/Na, DFT (DFT-D)). Atom color scheme is the same as in Figure 1.

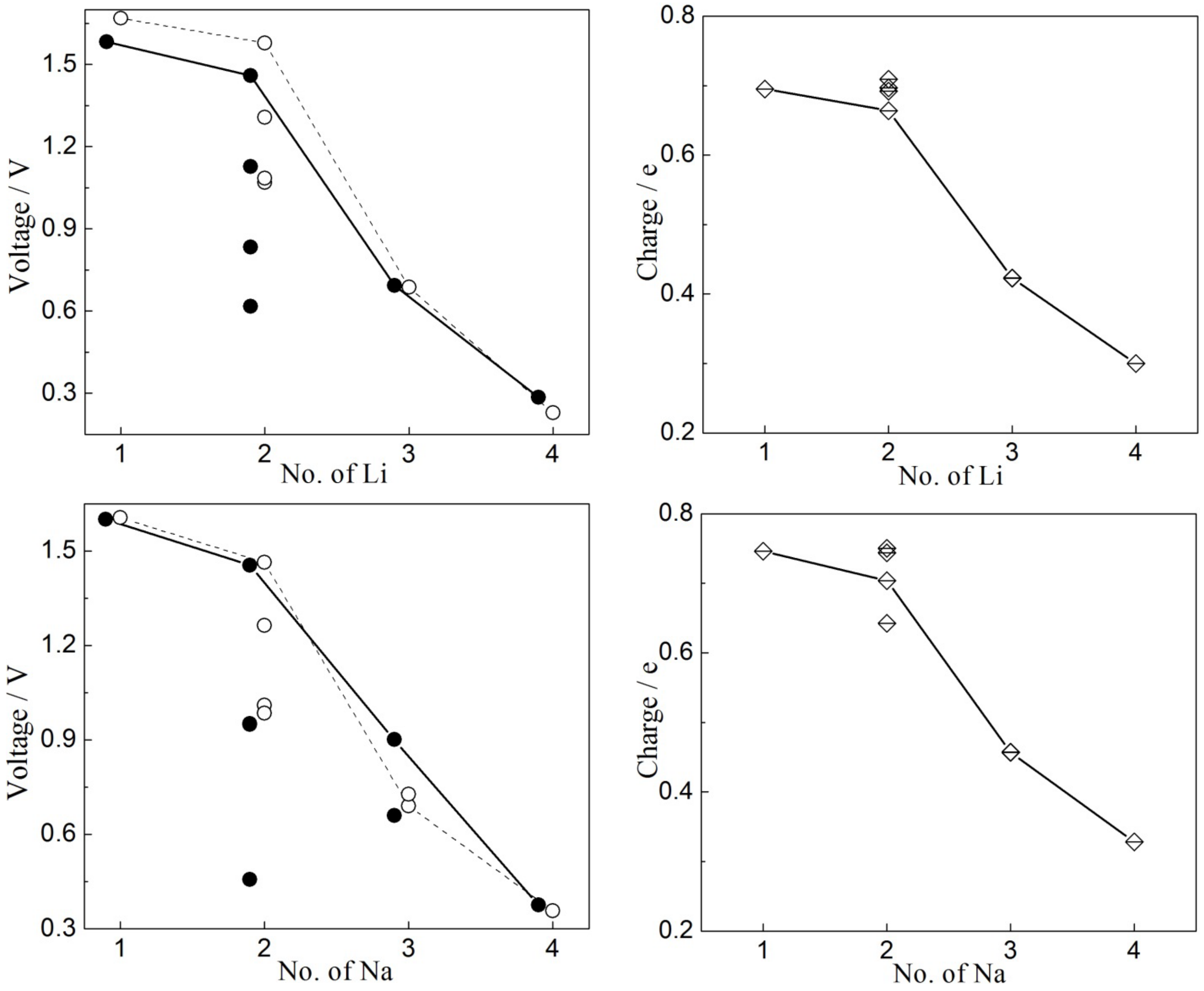

**Figure 4.** Left: DFT (filled circles and solid lines) and DFT-D (empty circles and dashed lines) values for voltages ($E_{coh}^{Li/Na}$-$E_b$) of different *n*Li/Na-TCNE complexes. Right: average charge donation (per Li/Na atom) in different *n*Li/Na-TCNE complexes. The lines connect lowest-energy configurations. Top panels are for Li and bottom for Na.

Li and Na adsorption configurations on TCNE-O-Al-graphene and their Li binding energies are shown in Figure 3. The number of configurations is higher than that for attachment to a free TCNE molecule due to the symmetry breaking by the presence of the substrate (Figure 2, right). Also, we were able to identify stable complexes with five Li/Na atoms, compared to a maximum of four for a free molecule. The values or $E_b$ are lower (more negative) compared to those with the free TCNE and should result in a higher voltage. In Figure 4 and 5, we plot the expected “voltage” profiles ($V = E_{coh}^{Li/Na} - E_b$) for Li/Na storage on TCNE and TCNE-O-Al-graphene. It can be seen that the inclusion of the Grimme potential changes the voltage by up to 0.3 V but does not change the curve qualitatively, so that both of DFT and DFT-D predict feasibility of Li/Na storage. Significantly, the computed voltage for the TCNE-O-Al-graphene system, at least for up to 3 Li/Na atoms per TCNE molecule, is higher than the reduction potential of common liquid organic electrolytes, such as $LiPF_6$ in EC:DEC/DMC at around 1.3 V [37] and $NaClO_4$ in EC:DMC at around 1.2 V [38] (vs. $Li/Li^+$ and $Na/Na^+$, respectively). This anode material could therefore avoid the formation of the SEI and associated kinetic hindrance and loss of Li/Na.

Also in Figure 4 and 5 (right) shown is the average charge donation per Li or Na atom from Li/Na to the molecular system, based on Mulliken charges. The charge donation is significant, reaching about half an electron in 1Li/Na-TCNE complexes. It, however, becomes progressively smaller as the number of Li/Na atoms attached to free TCNE increases. This decrease in charge donation also tracks well the dependence of $E_b$ or voltage on the number of attached atoms. Na donates more charge to TCNE than Li. On the other hand, in the TCNE-O-Al-graphene system, the average amount of charge donation per Li/Na atom changes little and remains at around half an electron even as the number of attached Li/Na atoms increases. This is due to charge exchange with the substrate, as can be seen from Figure 6. As a result of this ability to accommodate charge, the binding energies are stronger on the adsorbed system than on the free molecule and have a less steep dependence on the number of Li/Na atoms.

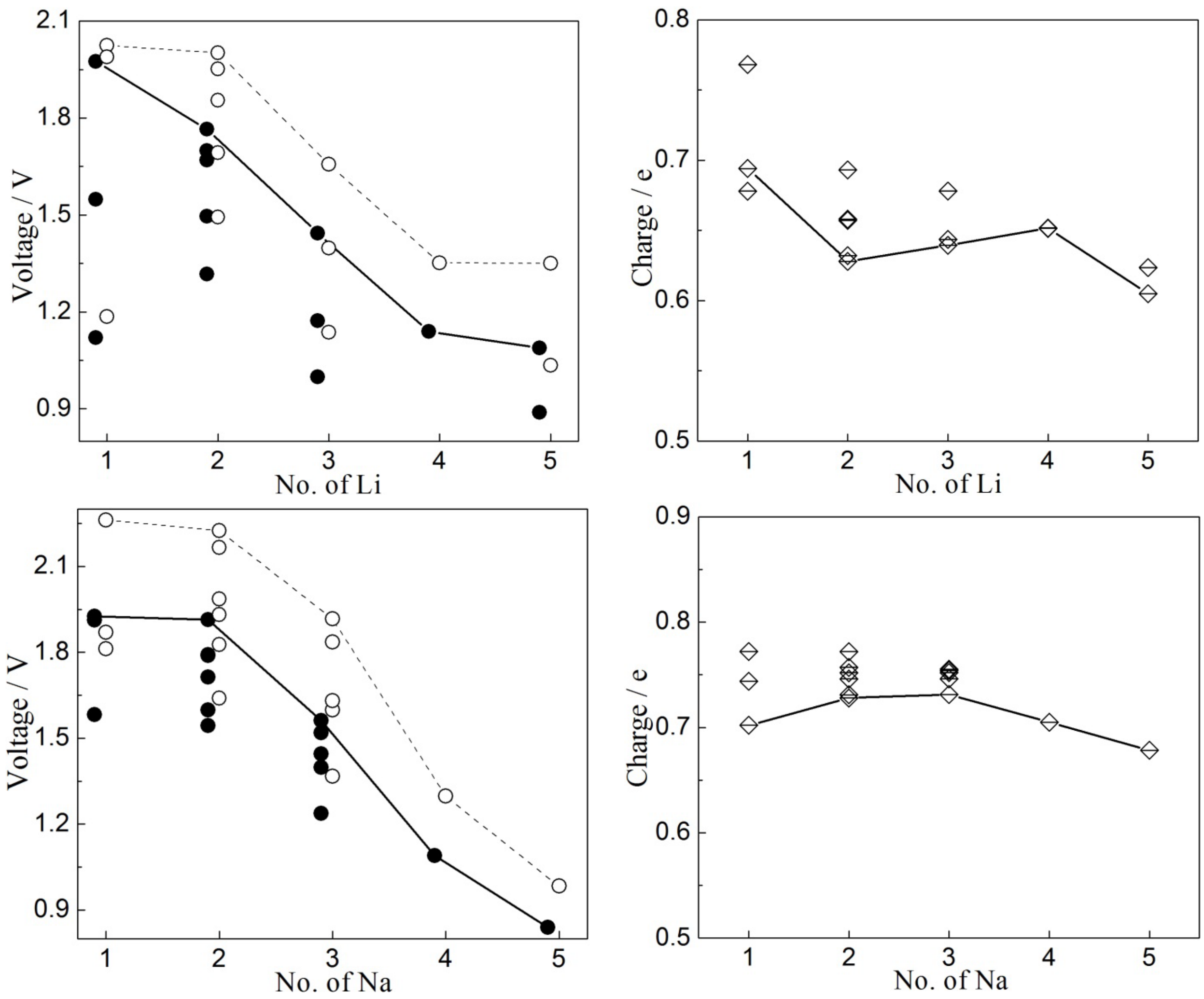

**Figure 5.** Left: DFT (filled circles and solid lines) and DFT-D (empty circles and dashed lines) values for voltages ($E_{coh}^{Li/Na}$-$E_b$) of different *n*Li/Na-TCNE-O-Al-graphene complexes. Right: average charge donation (per Li/Na atom) in different *n*Li/Na-TCNE-O-Al-graphene complexes. The lines connect lowest-energy configurations. Top panels are for Li and bottom for Na.

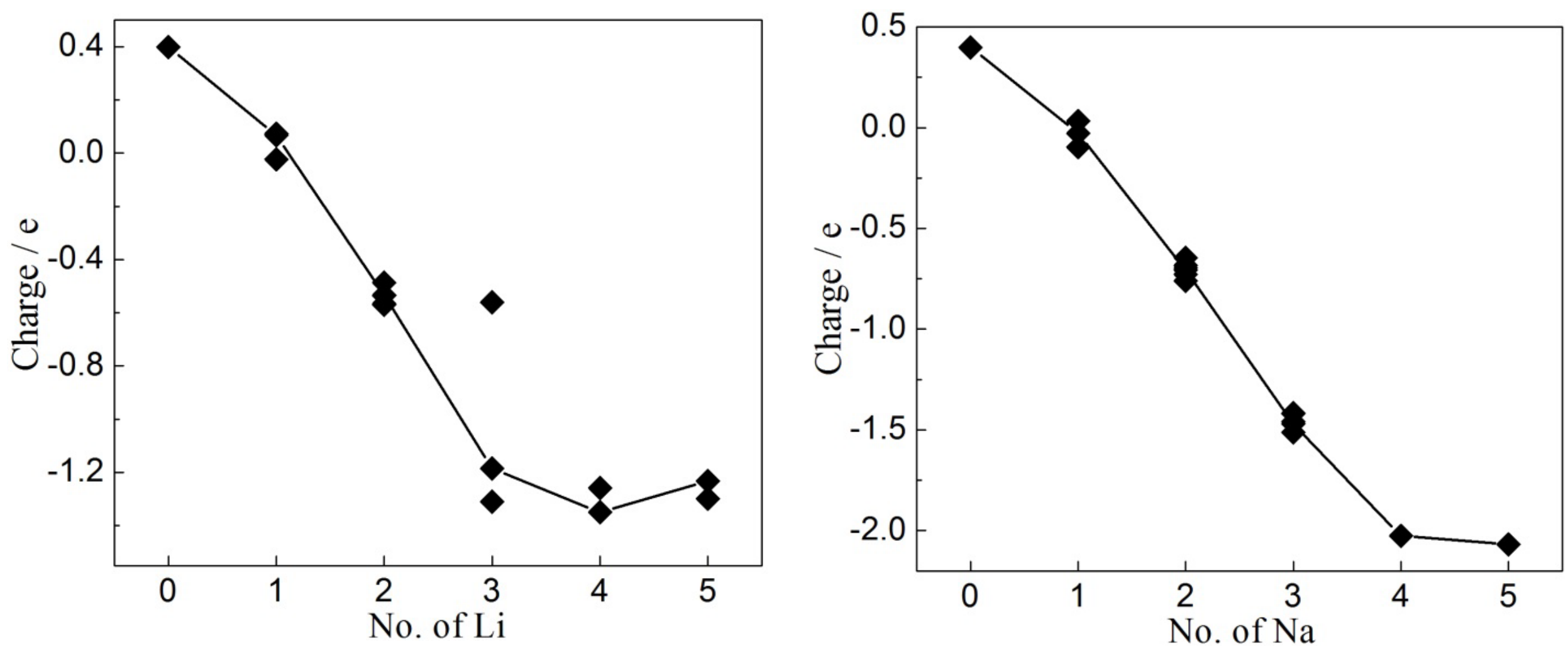

**Figure 6.** Sum of Mulliken charges on the Al-O-graphene moieties for different *n*Li/Na-TCNE-O-Al-graphene complexes. Left is for Li and right for Na. The lines connect lowest-energy configurations.

Figure 7 shows examples of charge density difference maps $\Delta\rho$ (Eq. 2), where the formation of Li/Na-TCNE bonds can be identified as accumulation of charge density between Li/Na and N atoms. In some configurations, Li or Na atoms are near each other. The analysis of $\Delta\rho$ allows one to identify bond formation also between Li/Na atoms (Figure 7). What's more, in some configurations with adsorbed TCNE, Li/Na atoms are in the proximity of the graphene sheet. The analysis of $\Delta\rho$ allows one to identify bond formation between Li/Na and N atoms via an accumulation of the electron density, but not between Li/Na and the substrate's C atoms (Figure 7). While some contribution to $E_b$ from Li-graphene interaction is possible, it is likely minor. This is also corroborated by the fact that $E_b$ is not necessarily stronger for configurations where Li atoms are near graphene (Figure 3).

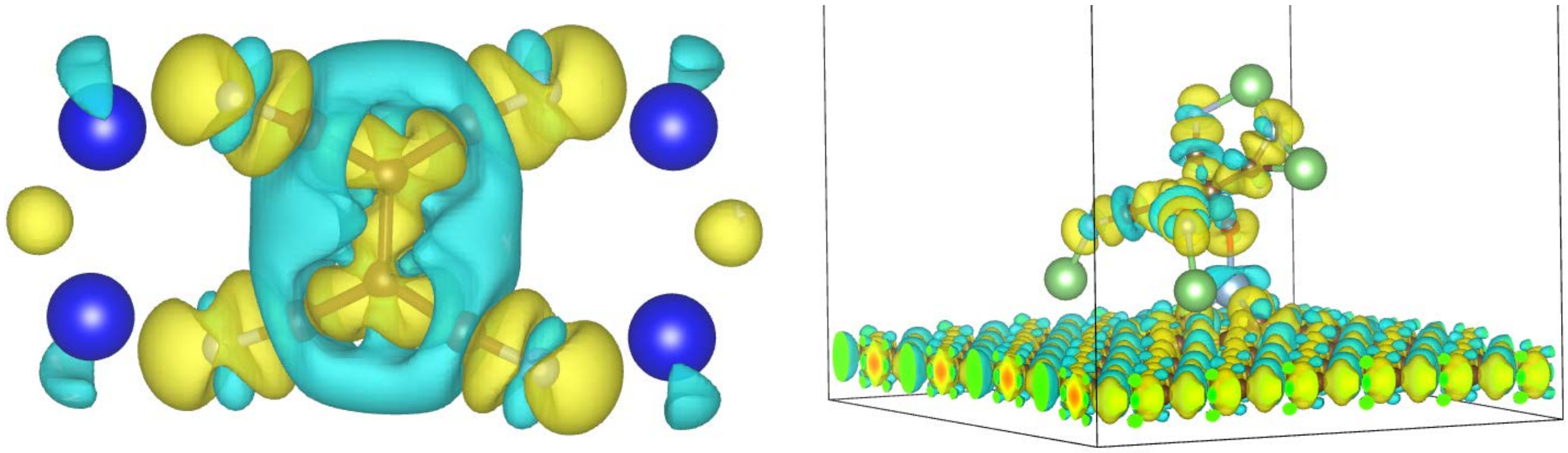

**Figure 7**. Examples of charge density difference (Eq. 2) for 4Li/Na-TCNE (left, cf. the image in Figure 1) and 4Li-TCNE-O-Al-grapnene (right, cf. the image in Figure 3). Yellow color corresponds to density accumulation, blue to depletion. Atom color scheme is the same as in Figure 1.

## 4. Conclusions

We studied computationally Li/Na storage at TCNE molecules as prospective organic anodes suitable for Li as well as Na ion batteries. We used both GGA DFT and DFT-D, and both methods lead to similar conclusions. Free molecules and molecules immobilized on doped graphene substrates were studied. A TCNE molecule can store up to 4 Li/Na atoms with binding energies larger than the cohesive energy of Li/Na metal, which would correspond to the specific capacity of about 840 mAh/g. TCNE adsorbs only weakly on ideal graphene. We identified an adsorption configuration on Al-doped graphene with strong chemisorption which preserves available all Li/Na attachment sites of TCNE. The molecule can therefore be anchored on a high surface area conducting substrate assuring the connectivity and stability of the electrode. Up to 5 Li/Na atoms can be stored on adsorbed TCNE with binding energies larger than the cohesive energy of Li/Na metal. The Li/Na-TCNE binding energy decreases with the number of attached atoms increasing, and this trend corresponds to the drop in the average charge donation from Li/Na to the molecule. In the Li/Na-TCNE-O-Al-graphene complexes, however, the average amount or charge donation to the molecule is little dependent on the number of attached Li/Na atoms, due to charge redistribution with the substrate. The binding energies are stronger on the adsorbed system than on the free molecule and have a less steep dependence on the number of Li/Na atoms. The voltage of such an anode molecule is predicted to be 0.89-1.98 (1.03-2.02) V for Li storage and 0.84-1.93 (0.98-2.26) V for Na storage with DFT (DFT-D), i.e. SEI formation could be avoided (if up to 3 Li/Na atoms per molecules are stored). Contrary to most inorganic electrode materials [11, 39-42], there is no significant difference either in specific capacity (per unit mass of material excluding Li/Na) nor in voltage between Li and Na storage, which makes this kind of approach very promising for post-Li storage. Even though the necessity of the substrate will decrease the specific capacity in a way dependent on the coverage, our calculations indicate that TCNE immobilized on a conducting graphene-based substrate could become an efficient anode material for organic Li and Na ion batteries.

## 5. Acknowledgements

This work was supported by Tier 1 AcRF grant R-265-000-494-112 by the Ministry of Education of Singapore. Y.Q.C. thanks Fleur Legrain, Department of Mechanical Engineering, NUS, for assistance with ab initio calculations.